\documentstyle[preprint,aps]{revtex}
\tighten
\input psfig.tex
\begin{document}
\draft

\title{Search for Kosterlitz-Thouless transition in a triangular Ising\\
antiferromagnet with further-neighbour ferromagnetic interactions}
\author{S. L. A. de Queiroz$^a$\cite{email} and Eytan Domany$^b$}
\address{
$^a$Instituto de F\'\i sica, Universidade Federal Fluminense,\\
Avenida Litor\^anea s/n,
Campus da Praia Vermelha,\\ 24210--340 Niter\'oi RJ, Brazil\\
$^b$Department of Physics of Complex Systems, Weizmann Institute of Science,\\
                  Rehovot 76100, Israel}
\date{\today}
\maketitle
\begin{abstract}
  We investigate an antiferromagnetic triangular Ising model with
anisotropic
ferromagnetic interactions between next-nearest neighbours,
originally proposed by Kitatani and Oguchi (J. Phys. Soc. Japan {\bf 57}, 1344
(1988)).
The phase diagram as a function of temperature and the ratio between
first- and second- neighbour interaction strengths is thoroughly examined.
We search for a Kosterlitz-Thouless transition to a state with algebraic decay
of correlations, calculating the correlation lengths on strips of width up to
15 sites by transfer-matrix methods. Phenomenological renormalization,
conformal invariance arguments, the Roomany-Wyld approximation and a direct
analysis of the scaled mass gaps are used.
Our results provide limited evidence that a  Kosterlitz-Thouless phase is
present. Alternative scenarios are discussed.
\end{abstract}

\pacs{PACS numbers: 75.30.Kz, 75.10.Hk, 05.50+q}
\narrowtext
\newpage

\section{ Introduction}

Some spin systems display an intermediate critical phase, which is
characterised by algebraic decay of correlations for a finite temperature
range,
and lies between the low-temperature ordered state and the high-temperature
paramagnetic regime.
This was first reported by Kosterlitz and Thouless (KT)~\cite{kt} in their
study
of the two-dimensional XY model, which exhibits long-range order only at $T=0$.
The KT transition from the paramagnetic to the critical phase is expected to
involve an exponential divergence of the correlation length~\cite{kt}.

 Later it was found that some two-dimensional systems with discrete spin
symmetry could accommodate critical phases as well.
The first systems to be discovered in this category were the $Z(N)$ models with
$N>4$\cite{jose,elitzur}.
 Jos\'e  $et\ al.$\cite{jose} found that, for the XY model with sixth-order
anisotropy,  the upper ($T_{1}$) and lower  ($T_{2}$) limits of the KT phase
are associated with the following values  for  the decay-of-correlations
exponent  $\eta (T)$:
$\eta (T_{1})=1/4$ and $\eta (T_{2})=1/9$.
This result has been used\cite{landau2,kitatani,miyashita} as a criterion
to establish the existence and limits of a KT phase in other systems, including
the one which concerns us in the present work.

Another class of spin systems with discrete symmetry displaying critical phases
is obtained from models with a macroscopically degenerate ground state, by
adding degeneracy-lifting
fields or interactions. Such is the case of the nearest-neighbour Ising
antiferromagnet on the triangular lattice. At $T=0$ this model has a finite
entropy per spin\cite{hout}, and algebraically decaying
correlations\cite{steph}. The behaviour of this system in a uniform magnetic
field and with second-neighbour ferromagnetic interactions was studied by
Nienhuis {\it et al.}\cite{nienhuis}. They predicted that, when the nearest
neighbour interaction, $K_{nn}$, is set to $-\infty$ and the reduced field
$H = B/T$ is small, the algebraic phase persists, until it is destroyed at
high fields. These predictions were verified subsequently\cite{blote,blote93}.
Other predictions, involving addition of finite (reduced) second-neighbour
ferromagnetic interactions, $K_{nnn}$, were made; it is expected that for large
enough $K_{nnn}$ the algebraic phase exists also off the $K_{nn} = -\infty$
line
({\it i.e.}, for finite $K_{nn}$); when $K_{nnn}$ is increased further,
however, a transition to a phase with long range order is expected. This is
consistent
with other results\cite{landau2,domany1} (see below).

The triangular Ising antiferromagnet with second-neighbour
ferromagnetic bonds was initially
proposed to model physical adsorption of gases on graphite\cite{campbell};
alternatively, it was used to describe the magnetic behaviour of CsCoCl$_3$
and CsCoBr$_3$\cite{mekata}.The model was
predicted\cite{domany1} to be in the universality class of the XY model with
sixth-order anisotropy\cite{jose}, and thus to exhibit a KT phase. Monte
Carlo results, together with the criterion of Jos\'e $et\ al.$, provided
numerical evidence for this\cite{landau2}.

A phenomenological renormalisation group (PRG) analysis, however, produced
clear signs of only one of the two transition temperatures associated with
the KT phase\cite{saito}. This discrepancy was ascribed to the small strip
widths used, namely 3 and 6. In order to be able to analyse larger strip
widths, Kitatani and Oguchi\cite{kitatani} introduced a slightly different
model which incorporates only four of the six second-neighbour interactions
per site. Each of the three second-neighbour sublattices  then forms a square
lattice, instead of the original triangular one.
This introduces spatial anisotropy, with consequences which we shall discuss
in what follows. In the original work of Ref.~\onlinecite{kitatani}, the
underlying hypothesis is
that, since the two-dimensional character
of the intrasublattice links was preserved, the qualitative behaviour ought
to be in the same universality class of the model with all six interactions
per site. Those authors managed to study strips of width up to 12 sites
with periodic boundary conditions (to which the Roomany-Wyld
approximation\cite{rw} for the $\beta$ function\cite{barber} was applied), and
17 with helical boundary conditions (for which the spin-spin correlation
functions were calculated directly and fitted to the criterion of
Jos\'e $et\ al.$).
While results from the former approach were deemed inconclusive, the authors
pointed out that the latter provided evidence of a KT phase.
Subsequent Monte Carlo work was found to be consistent with
this\cite{miyashita} by means
of a finite-size scaling and renormalisation-group analysis of the simulation
data, partly with help from the criterion of Jos\'e  $et\ al.$.
Very recently, free energy and magnetisation were calculated for this
model\cite{pajersky} by a combination of transfer-metrix and mean-field ideas.
The authors find an intermediate, incommensurate phase.
The KT transition, in this interpretation, would be commensurate-incommensurate
in character. An exact analysis of the ground-state of the model with
second-neigbour bonds along only {\it one} direction also shows evidence of a
critical phase\cite{noh}; when interactions along a second direction are
included, so as to make the model identical with the present one,
an exact treatment is no longer possible, even for the ground state. However,
Monte-Carlo and transfer matrix data also  point towards a
commensurate-incommensurate transition at zero temperature ($i.e.$ when the
first- neighbour couplings are set at $-\infty$) in this case, as the strength
of second-neigbour bonds varies\cite{nk2}. The latter results are not to be
directly compared to ours, as they pertain to the interacting domain-wall
transitions which take place between ground-state configurations.

Here we study the model proposed by Kitatani and Oguchi\cite{kitatani}
on strips of width 3, 6, 9, 12 and 15 with periodic boundary conditions. The
interpretation of PRG results in searching for a KT phase is discussed, and
these are shown to be of limited usefulness when considered on their own.
The Roomany-Wyld (RW) approximation\cite{rw} for the $\beta$
function\cite{barber} is implemented, for comparison
with the corresponding results of Kitatani and Oguchi; a direct
analysis of the ratio between the correlation lengths on strips of different
widths (which, for quantum systems, correspond to the scaled mass
gaps)\cite{rw,hamer,domany2} is given as well.
We discuss the bearing of conformal invariance theory on the comparison of our
results to previous ones. Finally we try to provide a consistent framework for
our findings.

\section{ Phase Diagram; Specific Heats}

The model is characterised by the nearest-neighbour interaction $J_{nn}<0$
and the second-neighbour coupling $J_{nnn}>0$, or alternatively by the ratios
of these to the temperature, respectively $K_{nn}=J_{nn}/T$ and
$K_{nnn}=J_{nnn}/T$. A useful parameter is the ratio
$R~\equiv~J_{nn}/J_{nnn}$. Kitatani and Oguchi\cite{kitatani}
restricted themselves to $R=-2.0$, whereas Miyashita {\it et
al}\cite{miyashita}
considered both $R=-2.0$ and $R=-5.0$. We span the region $-20.0\leq R\leq
0.0$.

A phase diagram which, though not completely consistent with all our findings,
appears to be quite plausible, is presented in Fig. \ref{fig:pdext}. Near the
origin there is a high-temperature paramagnetic phase. Above the solid line
we predict existence of a low-temperature phase with long-range order. The
shaded region in between is our candidate for a critical (or massless) KT
phase.
Fixed-$R$ lines, along which detailed studies were carried out, are also
indicated.

An examination of specific heats along lines of fixed $R$ unveils a two-peak
structure. This is especially apparent for
$R= -2.0$, as shown in Fig. \ref{fig:sph1} for $L = 3, \dots , 15$, for
comparison with Fig. 3 of Ref.\onlinecite{kitatani}, where $L=9$ is the largest
width used. However, when other values of $R < -2.0$ are investigated,  the
high-temperature peaks are displaced to ever higher temperatures and turn much
broader, though those at the low end remain sharp and essentially unmoved.
Fig. \ref{fig:sph2} illustrates this, with data for $L=15$ and $R= -2.0$,
$-5.0$ and $-20.0$ .
The specific heat peaks apparently do not diverge with increasing $L$. Hence
their positions do not identify transition points.  The double-peak structure
can be used only
as a qualitative indicator of resemblance to $Z(N)$ models with $N>4$, which
also exhibit such structure, and for which an intermediate massless phase is
known to exist\cite{jose,elitzur}. We shall not return to specific heats;
instead, our work is focused mainly on correlation lengths and quantities
derived therefrom, which will prove to be a safer ground for analysis.

\section{ Phenomenological Renormalization}

We begin by performing standard phenomenological renormalisation. The  PRG
procedure consists of searching, for fixed $R$, for the fixed point
$\{K_{nn}^{*},K_{nnn}^{*}\}$ of the implicit recursion
relation
\begin{equation}
{\xi_{L^{\prime}}(K'_{nn},K'_{nnn})\over
L^{\prime}}={\xi_{L}(K_{nn},K_{nnn})\over L} \ ,
\label{eq:1}
\end{equation}
where $\xi_{L}$ is the correlation length along a strip of width $L$. Usually
best
results are obtained for $L^{\prime}/L \rightarrow 1$; as in the present
case sublattice symmetry demands that these be multiples of 3, we use
$L^{\prime}=L-3$. The correlation length is given by
$\xi_{L}=1/(\zeta~\log~\lambda_{1}/|\lambda_{2}|)$, where
the geometric factor $\zeta$ is\cite{blote,blote93} $(2/\sqrt{3})$ for
$K_{nn}\not= 0$ and
$\sqrt{2}$ for$K_{nn}=0$; $\lambda_{1,2}$ are
the two largest eigenvalues (in absolute value) of the $2^{L}\times 2^{L}$
transfer matrix
between adjacent columns of a strip of width $L$. As $\lambda_{1}$ is
related to the strip free energy, it is always positive, while $\lambda_{2}$
in the present case is one of a pair of complex conjugate eigenvalues with
arguments $\pm 2\pi /3$ (owing to the three-sublattice structure). If all six
second-neighbour bonds were present\cite{saito}, the transfer matrix would
have to take into account adjacent $pairs$ of columns, raising its
dimension to $2^{2L}\times 2^{2L}$.
Omission of two out of the six next-nearest neighbour couplings is thus
technically worthwhile. One should, however, bear in mind the fact that, by
doing this, the six-fold  symmetry of the problem has been broken. Even though
the model still has a six-fold degenerate ground state, the energies associated
with domain walls between these phases depend now on the orientation of the
walls, and one has to work out the symmetry of this new problem. It is quite
clear that the criterion $\eta =1/9$ of Jos\'e {\it et al}\cite{jose} to locate
the low-temperature transition to a six-fold symmetric ordered phase should
not be naively used here.

For a system with a KT phase, one could expect Eq. \ref{eq:1} to exhibit
a line of fixed points
along a $finite$ temperature interval, corresponding to the extent of the
intermediate phase~\cite{barber}.
Finite systems should display a precursor for such behaviour, such as having
more than one solution to Eq. \ref{eq:1} at nearby points, in the region where
a KT phase is expected.
 This does not happen in the present case, although the direct analysis of the
scaled mass gaps (to be discussed below) shows that
there are extensive temperature ranges along which other indicators of a
massless phase occur.
For each given $R$ and $L^{\prime}/L$, we find only one temperature as a fixed
point of Eq. \ref{eq:1}. In Fig.~\ \ref{fig:pd} we
display the approximate critical lines given by the
fixed points of Eq. \ref{eq:1} on the $(K_{nn},K_{nnn})$ plane, for the several
values of $L^{\prime}/L$. Points on the extrapolated curve are obtained by
keeping $R$ constant and plotting the respective
fixed-point couplings (e.g. $K_{nnn}$) for $L^{\prime}/L$~=6/9, 9/12 and 12/15
against $L^{-\psi}$, where $\psi$ is chosen to give the
best straight-line fit through the three points, and is an estimate of the
corrections-to-scaling exponent~\cite{derrida}. We find that
$\psi$ falls roughly in the range $\sim 1$--$3$ wherever such extrapolation is
applicable (see below).

We have also applied domain-wall renormalisation group ideas\cite{mcmillan}
by reversing the sign of the interactions that cross a line along the
strip, which amounts to establishing antiperiodic boundary
conditions\cite{cardyap}. The difference between the corresponding free energy
and that for a strip with standard periodic boundary conditions gives the
domain-wall free energy, which vanishes at
criticality\cite{watson} and is the quantity to be scaled. Again, only one
fixed point was found for each $R$. The extrapolated phase diagram thus
obtained (not shown here) is roughly the same as that coming from
correlation-length scaling.
Consequently, we did not pursue this line further. It is worth noting, however,
that the finite-$L$ data approach the extrapolation from the high-temperature
side while those from standard PRG (see  Fig.~\ \ref{fig:pd}) come from below.
This is to be expected from  well-known duality relations
between domain-wall energy and inverse correlation length\cite{watson}.

For isotropic systems,
conformal invariance\cite{cardy} allows one to extract additional information
from strip scaling, via the relationship
between correlation-length amplitudes on a strip of width $L$
at criticality and $\eta$, the
decay-of-correlations exponent :
\begin{equation}
 \eta=L/\pi \xi (T_{c}) \ .
\label{eq:2}
\end{equation}

Here, anisotropy is introduced by the missing second-neighbour bonds. Critical
correlations are expected to decay with different prefactors in different
directions, though with the same exponent. In some cases this can be explicitly
dealt with\cite{nb,bpp,kp}, leading to a modified form for Eq. \ref{eq:2} which
involves knowledge of correlation lengths along two different
directions\cite{nb}. Unfortunately this it out of reach in the present model,
whose very introduction was in order to enable one to build large-$L$ transfer
matrices along the specific direction where the bonds are missing. For
$K_{nn}=0$ or $K_{nnn}=0$, isotropy is restored and Eq. \ref{eq:2} is valid,
with results to be described below.

Referring to Fig.~\ \ref{fig:pd},
the following comments are in order:

(1) On the vertical axis ($K_{nn}=0$) one has three superimposed,
disjoint, square lattices; thus the $L$ arising in the conformal-invariance
expression for $\eta$ is one-third of the number of sites in the column. The
corresponding critical quantities,
extrapolated as above, are (exact values quoted in parenthesis):
$K_{nnn}(0)=0.441\pm 0.001 (0.4407...)$; $y_{T}=1/\nu=1.00\pm 0.01 (1)$;
$\eta=0.250\pm 0.001 (1/4)$.

(2) The extrapolated boundary leaves the vertical axis tangentially with
a crossover exponent $\phi \simeq 1.55$, where
$|K_{nn}|\sim~(K_{nnn}(0)-K_{nnn})^{\phi}$. This can be compared with the
exact $\phi= 7/4$ obtained by Slotte and Hemmer\cite{slotte} for the
corresponding model with antiferromagnetic second-neighbour interactions.

(3) In the region between $R\simeq -0.01$ and $R\simeq -0.3$ the
approximate critical lines become very close and cross each other twice.
Thus, an extrapolation similar to that of Ref.\onlinecite{derrida} is not
feasible.

(4) As $R\rightarrow -\infty$ the critical lines become horizontal;
the extrapolated value is $K_{nnn}=0.313\pm 0.001$. Already at $R=-2.0$
one is within less than 1\% of that. Qualitatively, as $|K_{nn}|$ grows
the only ratio that matters is that between the two $finite$
energies present, namely $J_{nnn}$ and $T$. Such quantity is analogous to the
magnetic field--to--temperature ratio in the zero-temperature nearest-neighbour
triangular antiferromagnet\cite{blote,blote93}. In order to check our results
against those of Refs.\onlinecite{blote,blote93}, we set $K_{nnn}=0$ and
make $|K_{nn}|$ large. This must coincide with their zero-field limit.
Indeed we obtain $\eta$ = 0.4815, 0.4909, 0.4947, 0.4965 respectively for
$L$ = 6, 9, 12 and 15, results in excellent visual agreement with Fig. 2 of
Ref. \onlinecite{blote} and converging towards the exact value\cite{steph}
$\eta=1/2$.
\medskip

The question remains of the interpretation of the critical line given by PRG
in the context of KT transitions, where $two$ phase changes are expected
to occur.
Following previous work\cite{kitatani,miyashita}, we try to fit
our results to the criterion of Jos\'e $et\ al.$\cite{jose}. To this end
we examine the exponent $\eta$
on the critical line.
Bearing in mind the above remarks concerning anisotropy, the quantity
$L/\pi \xi (T_{c})$ does not in general correspond to $\eta$; however, for
large  $|K_{nn}|$ and finite $K_{nnn}$  one has essentially a triangular
antiferromagnet with weak anisotropy induced by the square-lattice
ferromagnetic bonds. In this region, corresponding to the horizontal section of
the critical
line in Fig.~\ \ref{fig:pd}, Eq. \ref{eq:2} is expected to become
asymptotically
true.
Fig.~\ \ref{fig:eta} shows $L/\pi \xi (T_{c})$ as a function of $K_{nn}$.
It can be seen that for large  $|K_{nn}|$,  $L/\pi \xi (T_{c})$  becomes
constant and
extrapolates to 0.25 from below. On the basis of this, one would tend to state
that the critical line of PRG marks the high temperature boundary $T_{1}$ of
the KT phase, at least in the region where the antiferromagnetic coupling is
strong. This, however, is inconsistent with previous
findings\cite{kitatani,miyashita}
(see below). Further, inspection of the $\xi_{L}/L \ {\rm vs.}\ T$ curves
for large $|R|$
 shows no evidence of non-monotonic behaviour;
in particular, no trend is found towards a second crossing
close to temperatures where conformal invariance would give $\eta$ near 1/9.
An example is shown in Fig. \ref{fig:eta5}; for $R= -5.0$, at the couplings for
which $L/\pi \xi =1/9$ we do not see any tendency for the different-$L$ lines
to cross, or even to approach one another.
We then consider other quantities,
in order to estimate bounds for a possible KT phase.

\section{ Roomany-Wyld Approximation}

To make contact with previous results, we investigate the Roomany-Wyld (RW)
approximation\cite{rw} for the $\beta$ function\cite{barber}, as done in
Ref.~\onlinecite{kitatani} for $R=-2.0$. This is given by
\begin{equation}
\beta^{RW}_{LL^{\prime}}(T) ={{\log (\xi_{L}/\xi_{L^{\prime}})/\log
(L/L^{\prime}) -1}\over \{\xi_{L}'
\xi_{L^{\prime}}'/(\xi_{L}\xi_{L^{\prime}})\}^{1/2}}  ,
\label{eq:3}
\end{equation}
where $\xi^{\prime}$ denotes a derivative of $\xi$ with respect to temperature.
Note that $\beta^{RW}_{LL^{\prime}}=0$ at the fixed point of the PRG
transformation
(Eq. \ref{eq:1}), and one would expect it to remain at zero for an extended
temperature
interval corresponding to a KT phase. This would be in contrast with the
monotonic behaviour exhibited in systems with an ordinary
transition~\cite{barber}. In Fig.~\ \ref{fig:beta} we display
$\beta^{RW}_{LL^{\prime}}$ against
temperature for $R=-2.0$ and $L^{\prime}/L$=3/6, 6/9, 9/12 (already obtained
in Ref.~\onlinecite{kitatani}) and 12/15. Although the 3/6 curve is almost
featureless, a trend arises towards flattening as $L$
increases, especially when the 12/15
data are considered. The range of temperatures for which one would, from visual
inspection,
expect the $L \rightarrow \infty$ curves to touch the axis, is in broad
agreement with the approximate boundaries of
the KT phase as obtained by previous authors~\cite{kitatani,miyashita}.

 Fig.~\ \ref{fig:beta2} shows $\beta^{RW}_{LL^{\prime}}$ against temperature
for the largest strip width $L=15$ ($L^{\prime} =12$) and several values of
$R$. It becomes apparent that the trend towards flattening noticed earlier
turns even
weaker, on either side of $R=-2.0$.
As the RW approximation produces clear-cut plateaus in other cases where a
critical phase is present (see $e.g.$, Fig.6 of ref.\onlinecite{rw}), one is
led to concur with Kitatani and Oguchi's conclusion, by deeming the present
results inconclusive for general $R$.

\section{ Scaled Mass Gaps}

We then consider the ratio
\begin{equation}
{\cal Q}\equiv~(\xi_{L}(T)/L)/(\xi_{L^{\prime}}(T)/L^{\prime})\ .
\label{eq:4}
\end{equation}
This quantity was introduced in the study of the scaled mass gaps of quantum
systems~\cite{rw,hamer}, and was used in Ref.~\onlinecite{domany2} in the study
of anisotropic Ising triangular antiferromagnets in an external field, which
exhibit features  similar to those under examination here.
At the fixed point of PRG, ${\cal Q}=1$ and one expects it to remain
close to this value for an extended temperature interval if a critical phase is
present. This is numerically verified to a good extent in several
cases\cite{rw,hamer,domany2}.In Figures \ref{fig:rds2} and \ref{fig:rds5}
we show  $\cal Q$ respectively for $R=-2.0$ and $-5.0$, the same values
considered in Ref. \onlinecite{miyashita}, and several pairs of strip widths.
Figure~\ \ref{fig:rds} displays plots of $\cal Q$ against temperature for
$L=15$ and $L^{\prime} =12$, and $R=-0.5$, $-2.0$, $-5.0$ and $-20.0$.

Note that the fractional deviation of $\cal Q$ from
1 is much smaller than the corresponding shift of $\beta^{RW}_{LL^{\prime}}$
from zero
in Fig.~\ref{fig:beta2}. As $|R|$ grows, the extent of the flattening
region increases. For $R=-20.0$ and $L=15$, $\cal Q$ rises quite steeply from
0.95 at $T/J_{nnn}=2.4$ to 1.0
at $T/J_{nnn}\simeq 2.96$, the fixed point of PRG, and then remains below 1.005
up to $T/J_{nnn}=10.0$.

The data from $\cal Q$ indicate
that, at least for $R< -2.0$, there is a signature consistent with the presence
of a KT phase at temperatures above that of the fixed
point of PRG. This would support the results of
Refs.~\onlinecite{kitatani,miyashita}. Visual
extrapolation  of the temperature range for which $\cal Q$ approaches 1  for
large $L$ would give the upper limits  $T_1/J_{nnn} \simeq 4.0$ and $\simeq
5.2$  respectively for $R= -2.0$ and $-5.0$. These compare well with the
corresponding Monte Carlo results\cite{miyashita} $T_1/J_{nnn} \simeq 3.8$ and
$\simeq 5.0$ .
On the other hand, our evidence strongly suggests that the lower limit $T_2$
must be the fixed point of PRG, whose extrapolated location is at
 $T_2/J_{nnn} \simeq 3.19$, both for $R= -2.0$ and $-5.0$ . This value is
definitely higher than those quoted by Miyashita {\it et al}, resp. $\sim 2.6$
and $\sim 2.75$.

 For small $|R|$ the signature referred to above tends to disappear.
If the KT phase is present at all,
it must vanish as $R \rightarrow 0$, since $at$ $R=0$ one has a simple
structure of three independent,
superimposed, ferromagnetic square lattices. In this context, one cannot
rule out the possible
existence of a multicritical point at some $R_{t}\not= 0$, such that
for $|R|<|R_{t}|$ there would be only an ordinary transition. However, we
have not investigated this aspect in detail.

The trends arising from the examination of $\beta^{RW}$ and, especially,
 $\cal Q$ are difficult to reconcile with the results for $\eta$  coming from
conformal invariance
(though the latter are, admittedly, valid only for large $|R|$ when anisotropy
becomes weak).
 One would expect $\eta \simeq 1/4$ at the high-temperature
end of the massless phase, and $\eta < 1/4$ at low temperatures still within
the critical phase. This, however, is not seen in our results (recall, {\it
e.g.} Fig. \ref{fig:eta5}).

\section{ Discussion and Conclusions}

We have analysed
four
quantities for a wide range of values of the first-to-second-neighbour
interaction ratio $R$ in the Kitatani-Oguchi model:
$(a)$ the location of the fixed point of PRG; $(b)$ the critical exponent
$\eta$, as given by conformal invariance; $(c)$ the $\beta$ function in
the Roomany-Wyld approximation; $(d)$ the scaled mass gap ${\cal Q}$.
Overall, one can say that $(c)$
does not provide conclusive
evidence as regards the existence of a KT phase.
On the other hand, the positions of the seemingly clear-cut plateaus arising
from $(d)$, above the fixed point of PRG given by $(a)$, do indicate existence
of a massless phase; however the values of $\eta$ obtained in this region, from
$(b)$, are inconsistent with the standard expectations.

One possibility is that the model of Kitatani and Oguchi does $not$ display a
KT phase at all, owing to the fact (discussed above) that it is not six-fold
symmetric. In
this scenario, the transformation of the second-neighbour sublattices from
triangular to square via the deletion of the appropriate bonds would be
relevant in the renormalisation-group sense. For example, the work of
Jos\'e $et\ al$~\cite{jose} indicates that, for the $XY$ model with
fourth-class anisotropy one has only one transition, with $\eta=1/4$.

Although this is an appealing conjecture, we
feel that our data from $(d)$ for large $|R|$ definitely show a trend
towards the onset of a critical phase. Based on this, we propose the
approximate phase diagram shown in Fig. \ref{fig:pdext}, where the boundaries
of the shaded area are obtained from extrapolation of the flat sections of the
scaled mass gap
diagrams.

It is worth recalling
that other generalisations of the triangular Ising antiferromagnet also display
critical phases, without obeying the criterion of Jos\'e {\it et al}. We refer
specifically to the model with second-neighbour ferromagnetic bonds along only
one lattice direction\cite{noh}, and to one with continuous degrees of
freedom superimposed on the original Ising ones\cite{chandra}. For the present
model, we recall the zero-temperature transitions between domain
states\cite{nk2}, and single out the Monte-Carlo renormalisation group results
of Miyashita {\it et al}, summarised in Fig. 6 of Ref. \onlinecite{miyashita},
which do point towards the existence of a critical phase, and were
obtained without recourse to the criterion of Jos\'e {\it et al}. It would seem
that this latter criterion is valid only for the model for which it was
originally formulated\cite{jose} and its closest relatives\cite{landau2} with
full six-fold symmetry.

\acknowledgements

The authors would like to thank the Department of Theoretical Physics
at Oxford, where this  work was initiated, for the hospitality.
SLAdQ thanks D. P. Landau, C. Vanderzande. M. Henkel and J. Cardy for useful
conversations, and Departamento de F\'\i sica, PUC/RJ for use of their
computational facilities. Research of SLAdQ is partially supported by the
Brazilian agencies Minist\'erio da Ci\^encia e Tecnologia, Conselho Nacional de
Desenvolvimento Cient\'\i fico e Tecnol\'ogico and Coordena\c c\~ao de
Aperfei\c coamento de Pessoal de Ensino Superior. ED acknowledges partial
support of the
US-Israel Binational Science Foundation (BSF).


\newpage
\begin{figure}
\caption{
Qualitative phase diagram. Heavy line is extrapolated from PRG (see Fig. 4).
Shaded area gives massless phase, as obtained from scaled
mass gaps. Constant-$R$ lines are shown, along which detailed studies were
carried out. $T$ marks (conjectured) multicritical point (see text).
\label{fig:pdext}}
\end{figure}
\begin{figure}
\caption{
Specific heat against temperature for  $R$= $-2.0$. Bottom to top: $L$= 3, 6,
9, 12, 15 .
\label{fig:sph1}}
\end{figure}
\begin{figure}
\caption{
Specific heat against temperature for $L$=15 and $R$= $-2.0$, $-5.0$ and
$-20.0$ .
\label{fig:sph2}}
\end{figure}
\begin{figure}
\caption{Phase diagram in $K_{nn}$--$K_{nnn}$ space, as given by the fixed
points of
PRG. The heavy line is an extrapolation from the 6/9, 9/12 and 12/15 curves.
Some constant-$R$ lines are shown.}
\label{fig:pd}
\end{figure}
\begin{figure}
\caption{
$L/\pi \xi_L$
 along the critical curves of PRG.
The heavy line is an extrapolation from the 6/9, 9/12 and 12/15 curves.
\label{fig:eta}}
\end{figure}
\begin{figure}
\caption{
$L/\pi \xi_L$ against temperature for $R= -5.0$ . Horizontal lines at 1/9 and
1/4 mark the values of $\eta$ at the boundaries of the KT phase for the XY
model with sixth-order anisotropy.
\label{fig:eta5}}
\end{figure}
\begin{figure}
\caption{
Roomany-Wyld approximant $\beta^{RW}_{LL^{\prime}}$ against temperature for
$R= -2.0$, and $L/L^{\prime}$=3/6, 6/9, 9/12 and 12/15 .
\label{fig:beta}}
\end{figure}
\begin{figure}
\caption{
Roomany-Wyld approximant $\beta^{RW}_{LL^{\prime}}$ against temperature for
$L/L^{\prime}$=12/15 and $R$= $-0.5$, $-2.0$, $-5.0$ and $-20.0$ .
\label{fig:beta2}}
\end{figure}
\begin{figure}
\caption{
${\cal Q}\equiv~(\xi_{L}(T)/L)/(\xi_{L^{\prime}}(T)/L^{\prime})$ against
temperature for $R$= $-2.0$  and $L/L^{\prime}$=3/6, 6/9, 9/12 and 12/15.
\label{fig:rds2}}
\end{figure}\begin{figure}
\caption{
${\cal Q}\equiv~(\xi_{L}(T)/L)/(\xi_{L^{\prime}}(T)/L^{\prime})$ against
temperature for $R$= $-5.0$ and $L/L^{\prime}$=3/6, 6/9, 9/12 and 12/15 .
\label{fig:rds5}}
\end{figure}\begin{figure}
\caption{
${\cal Q}\equiv~(\xi_{L}(T)/L)/(\xi_{L^{\prime}}(T)/L^{\prime})$ against
temperature for $L/L^{\prime}$=12/15 and $R$= $-0.5$, $-2.0$, $-5.0$ and
$-20.0$ .
\label{fig:rds}}
\end{figure}
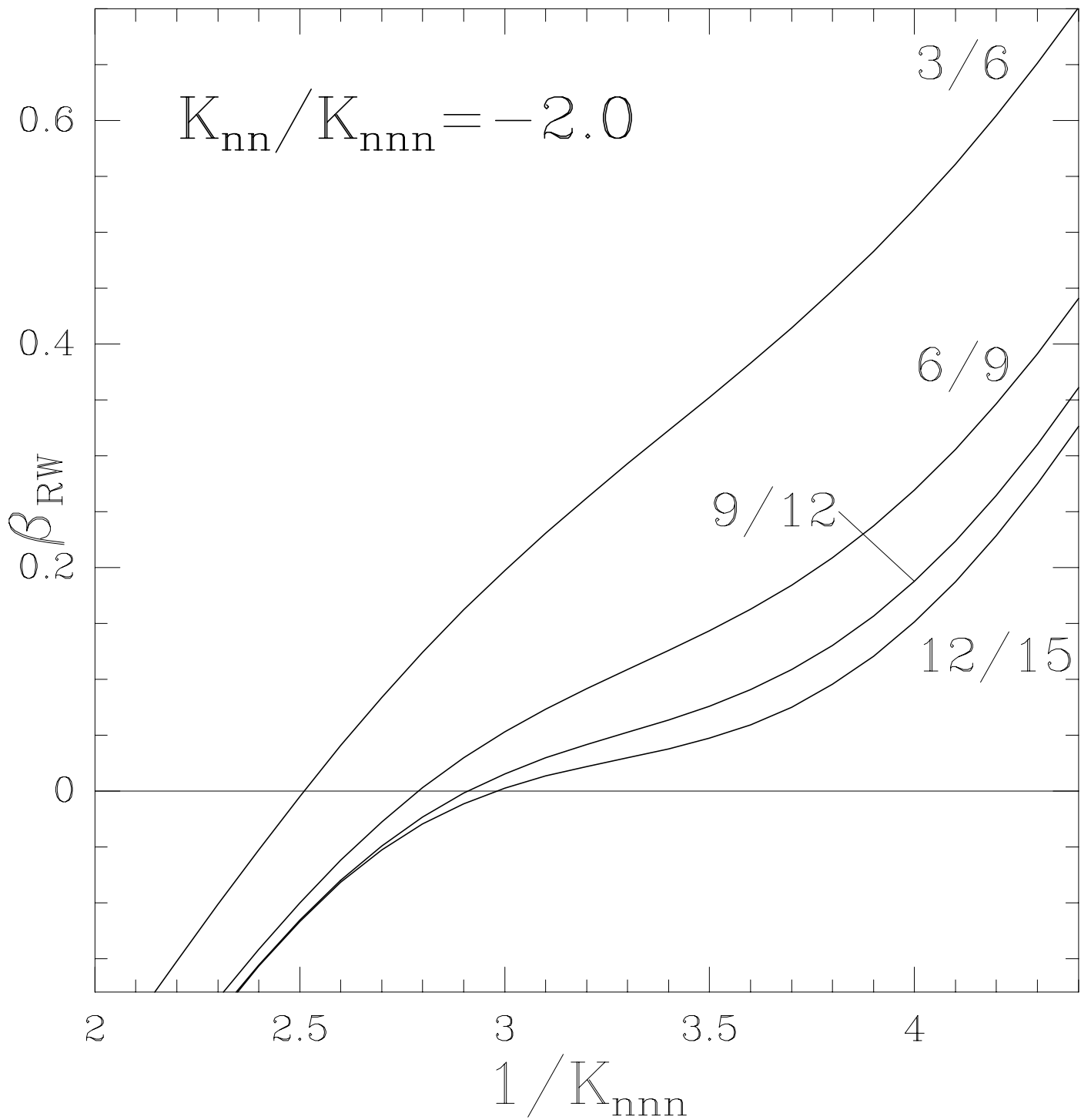
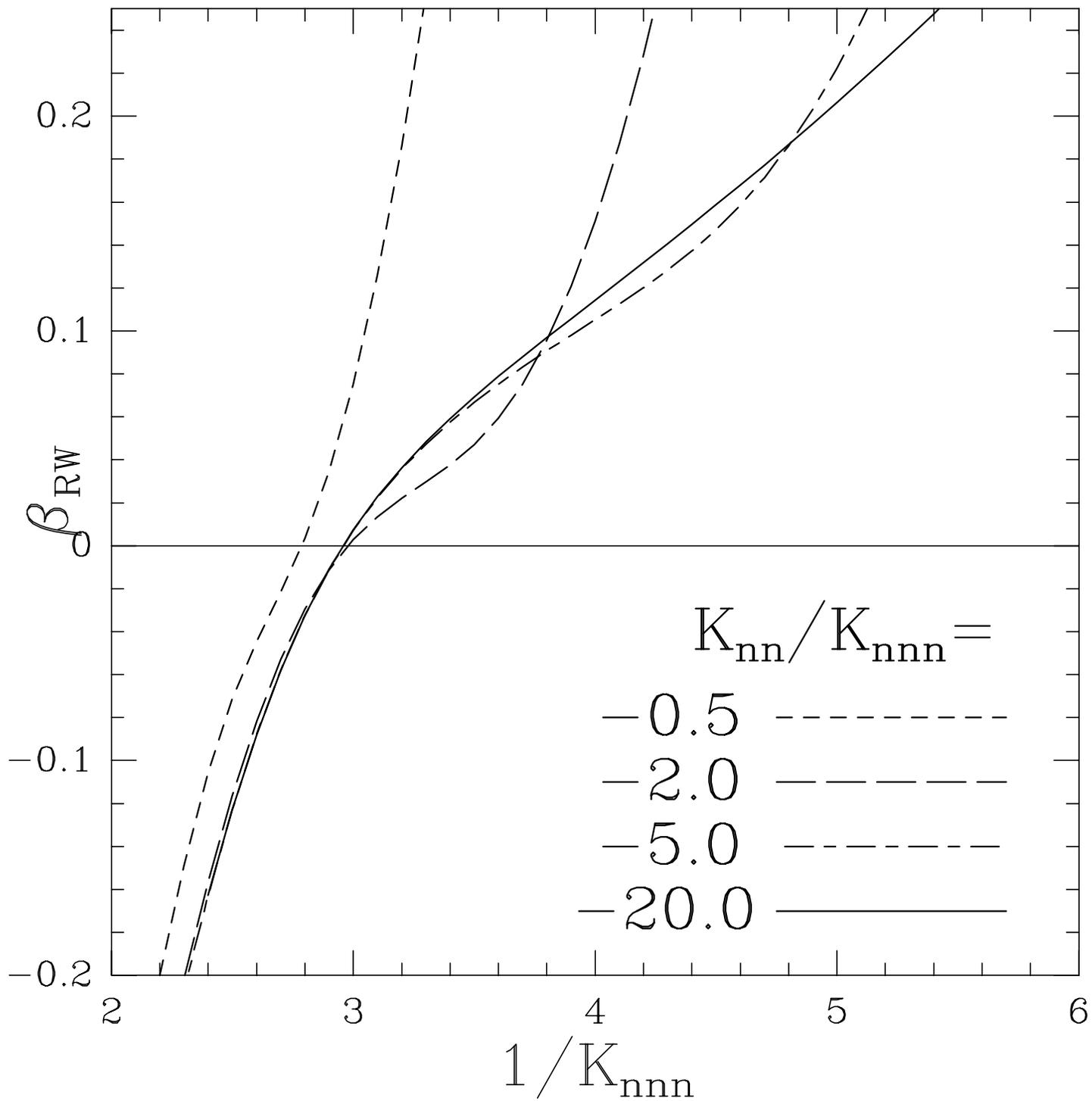
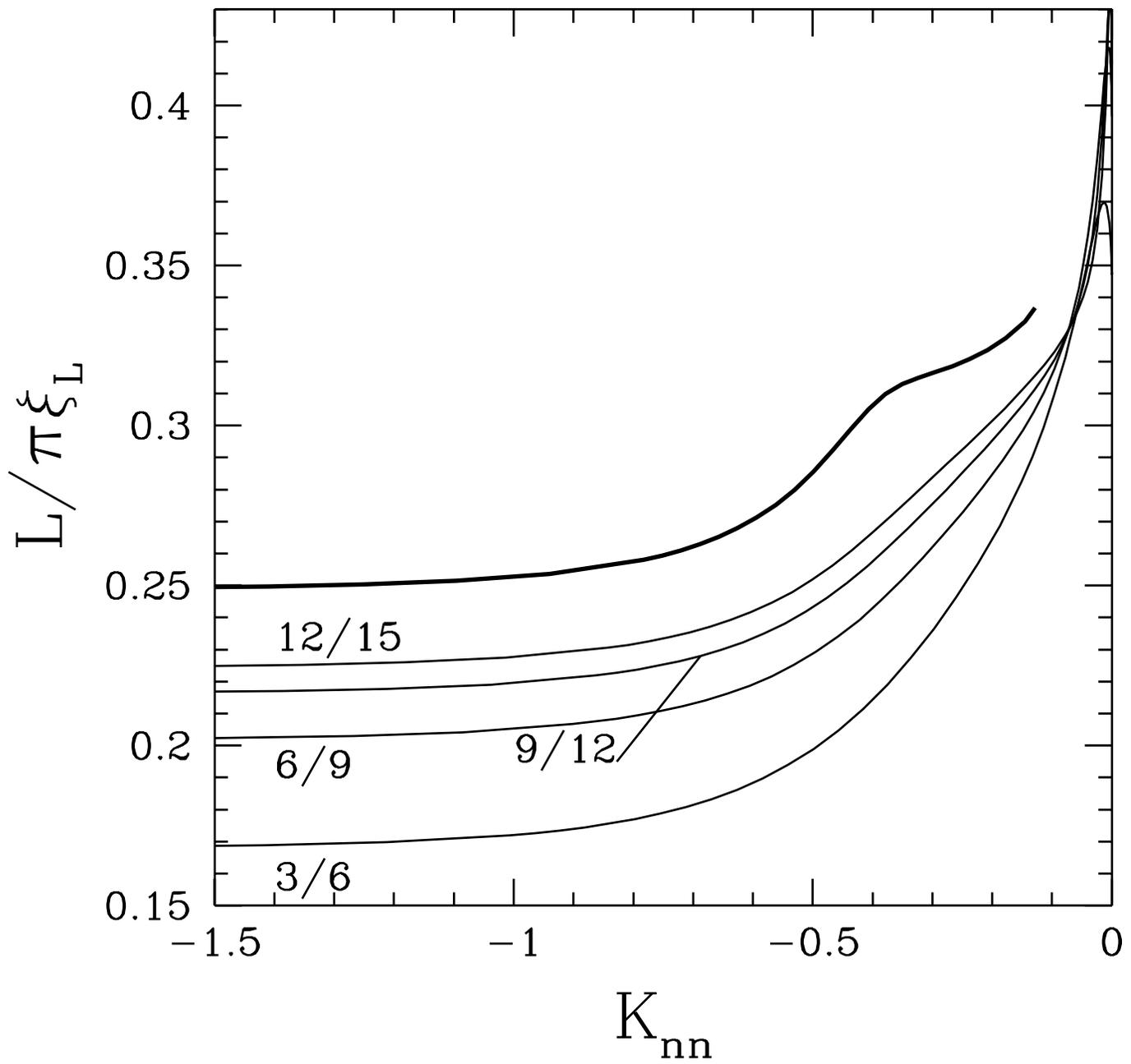
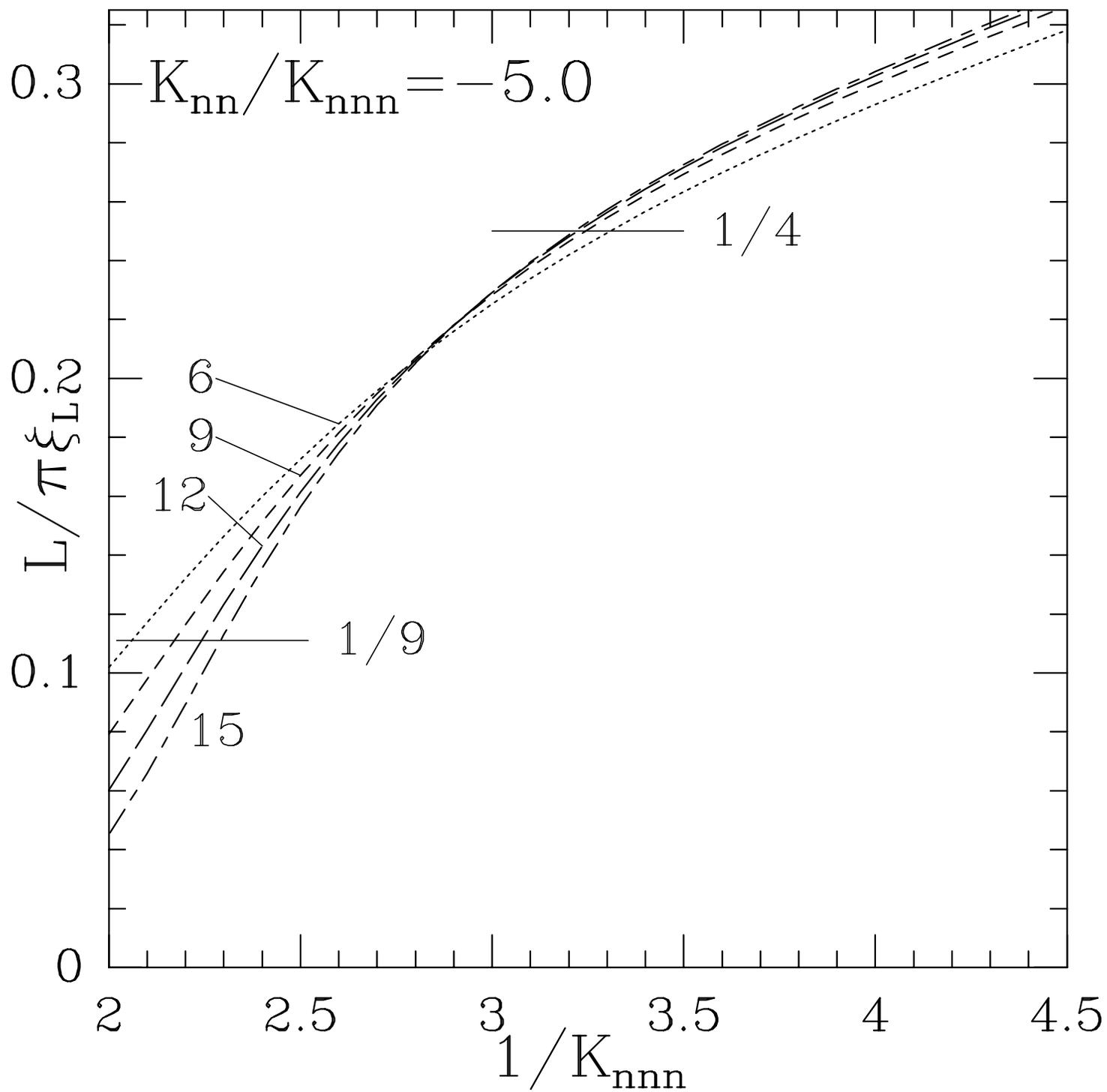
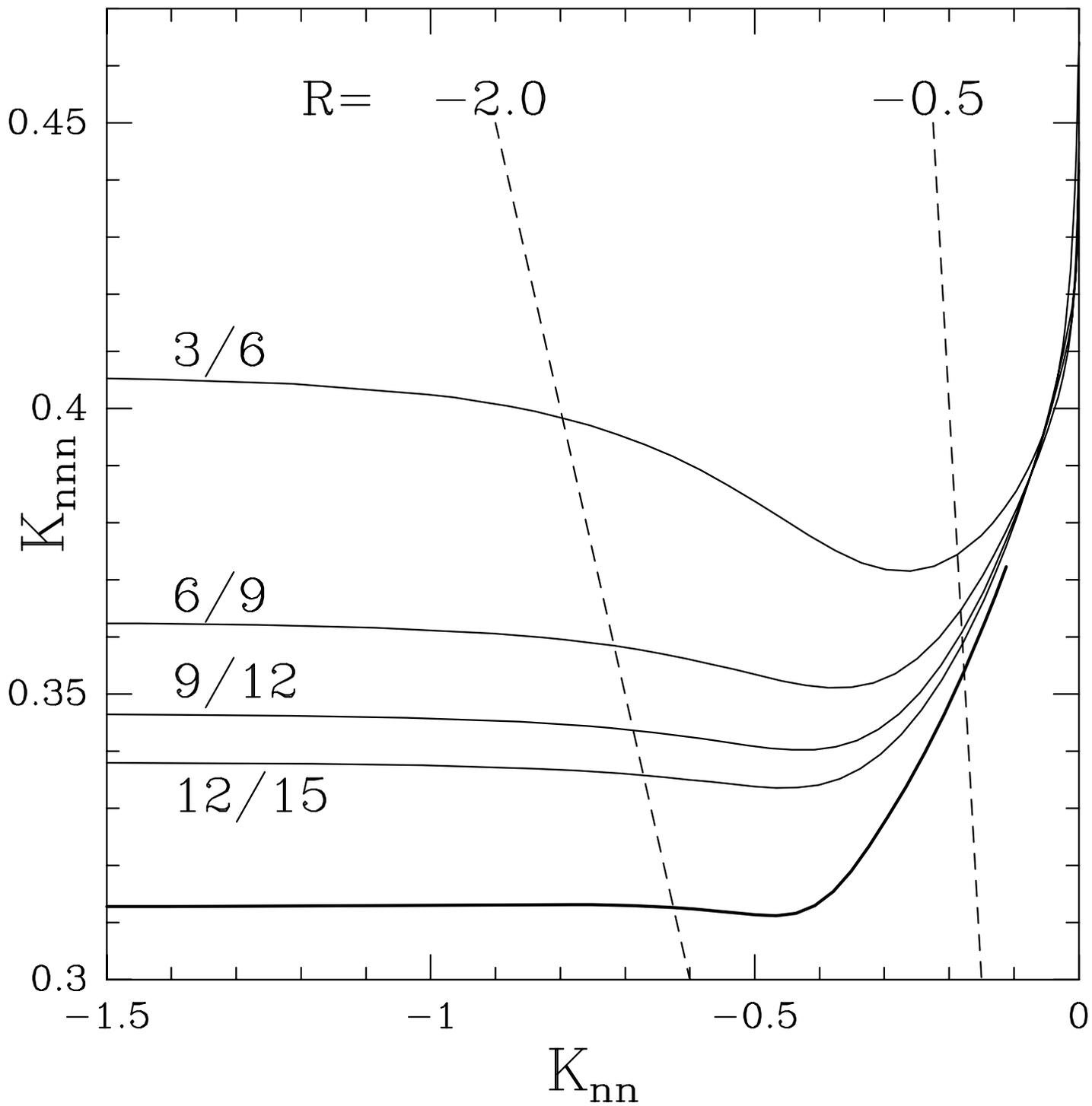
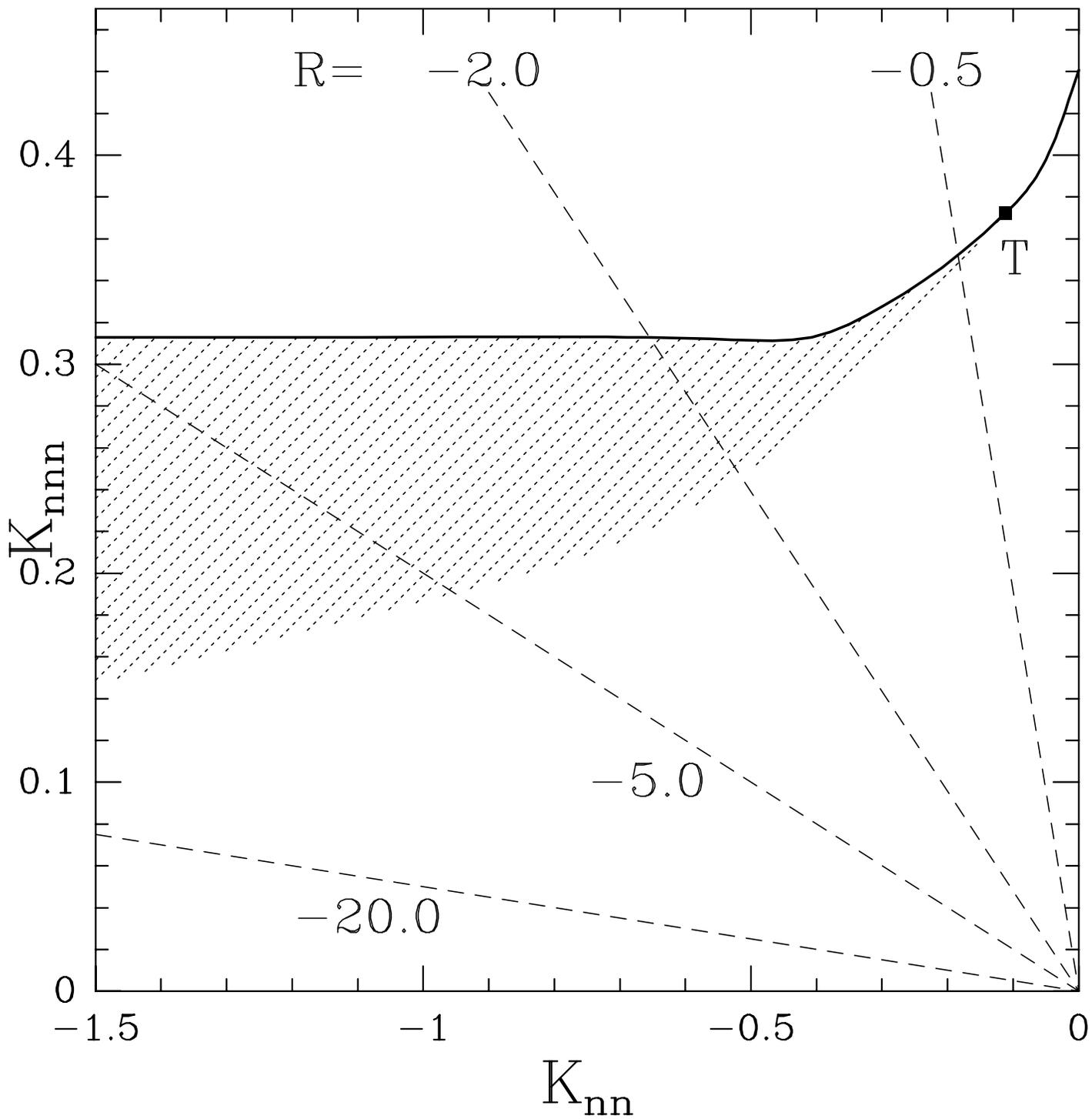
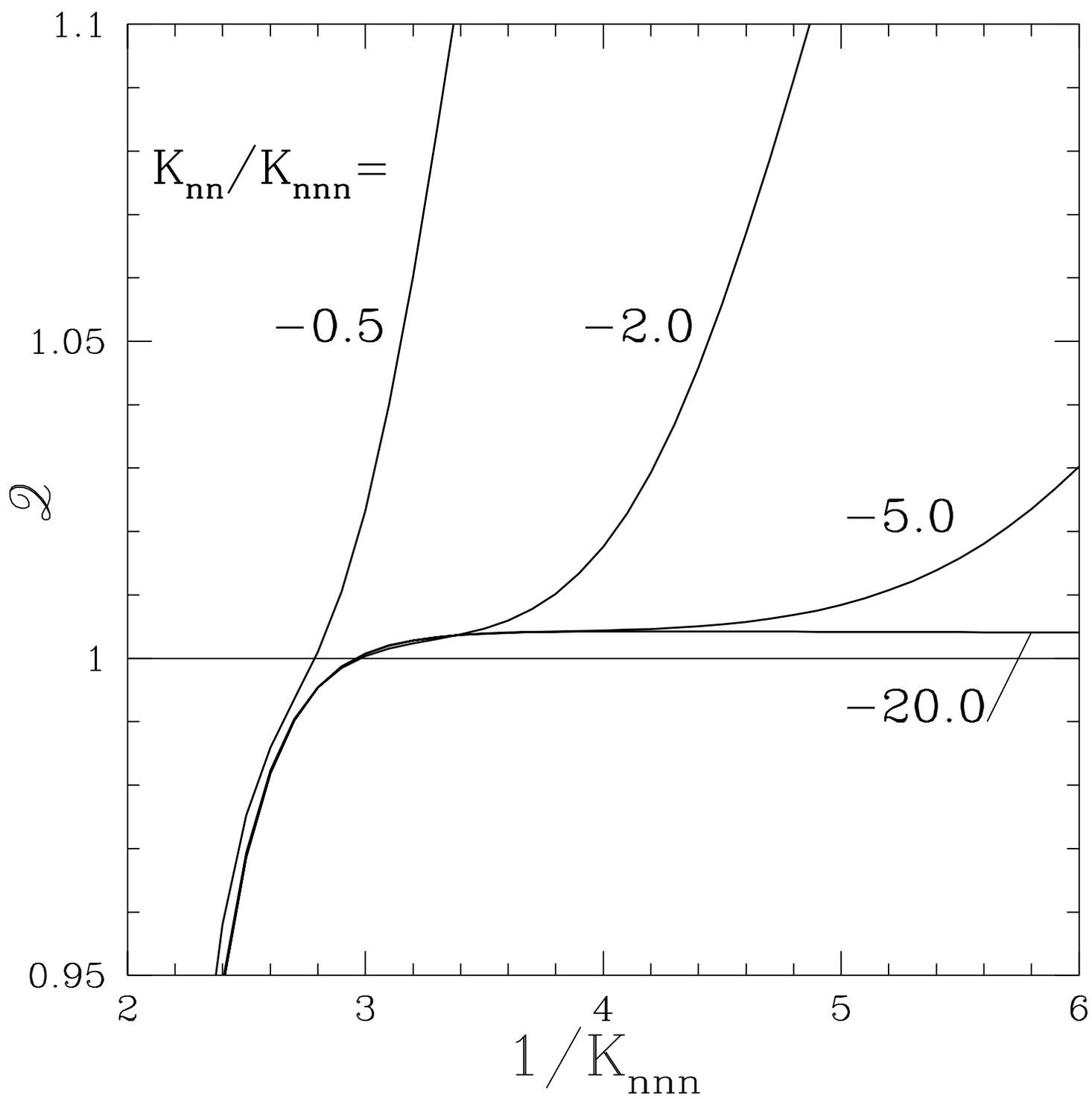
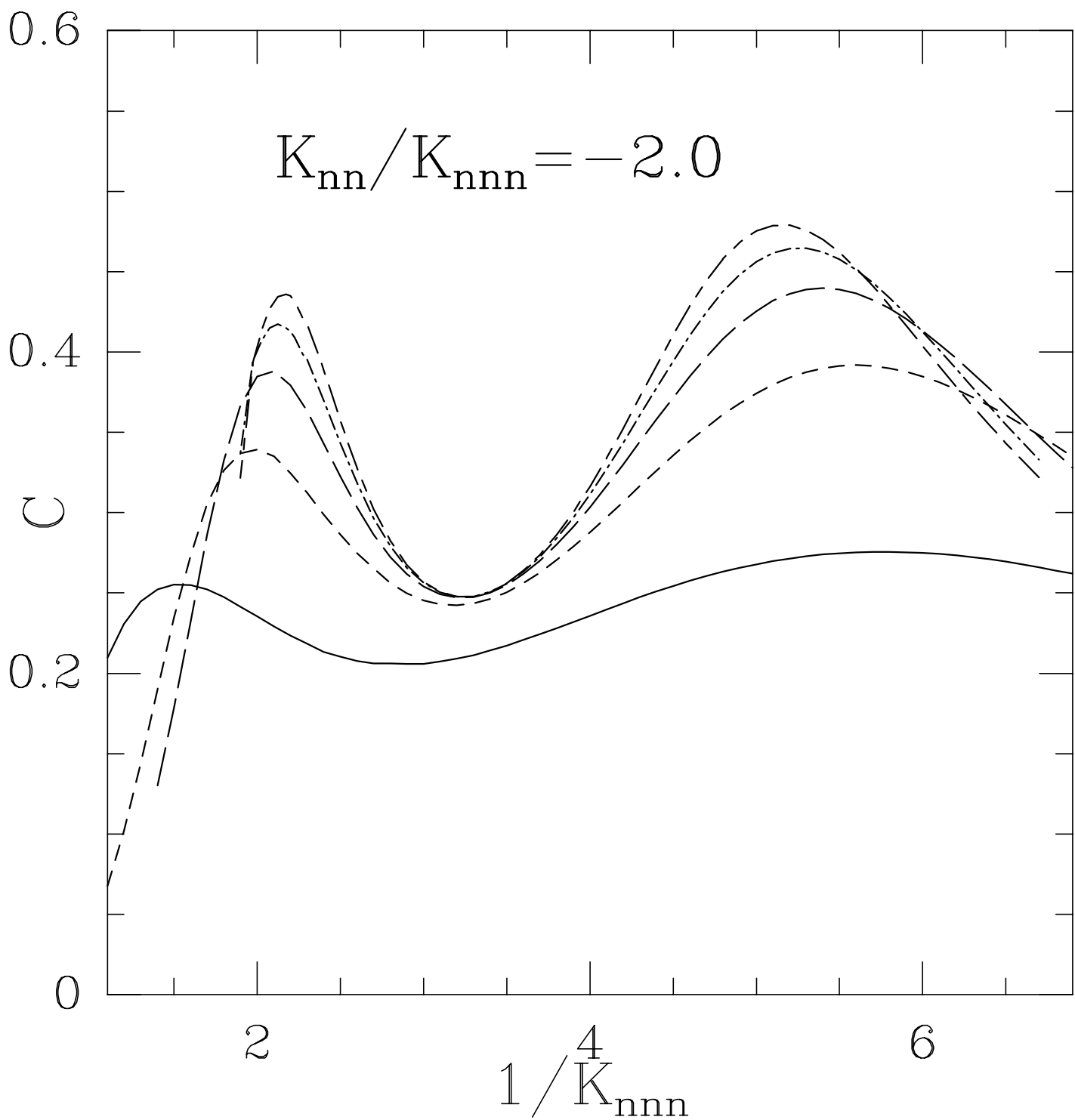
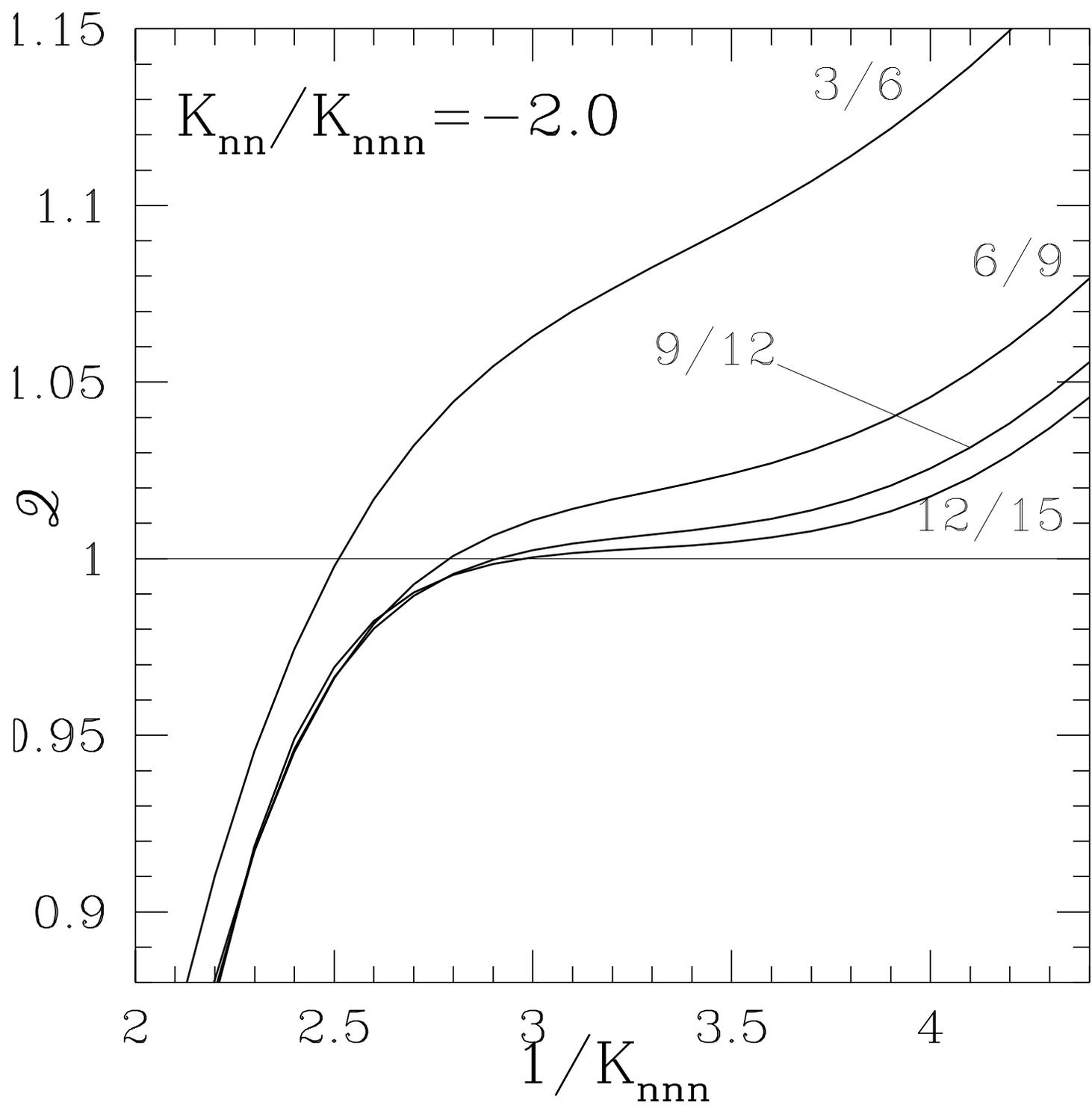
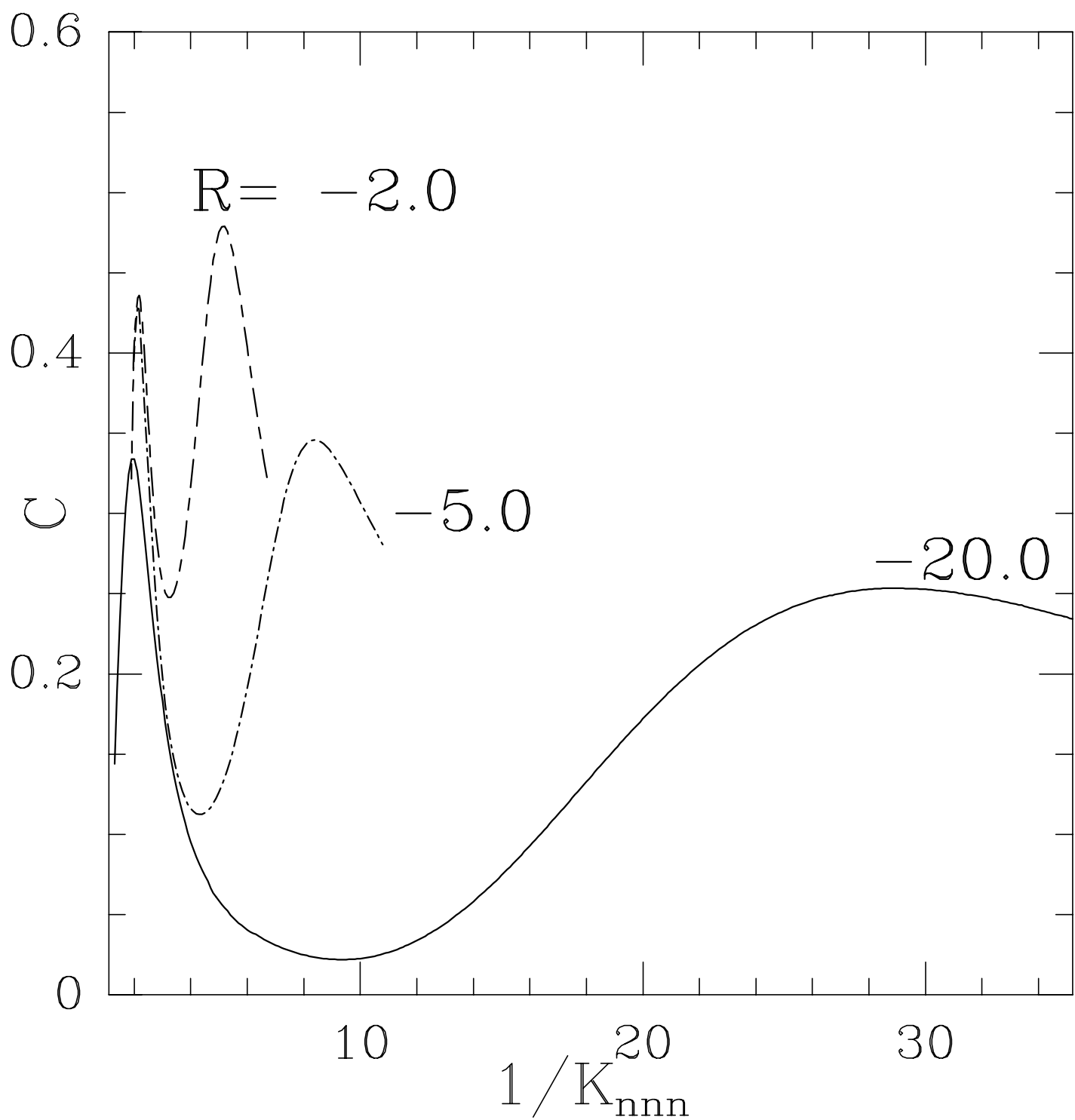
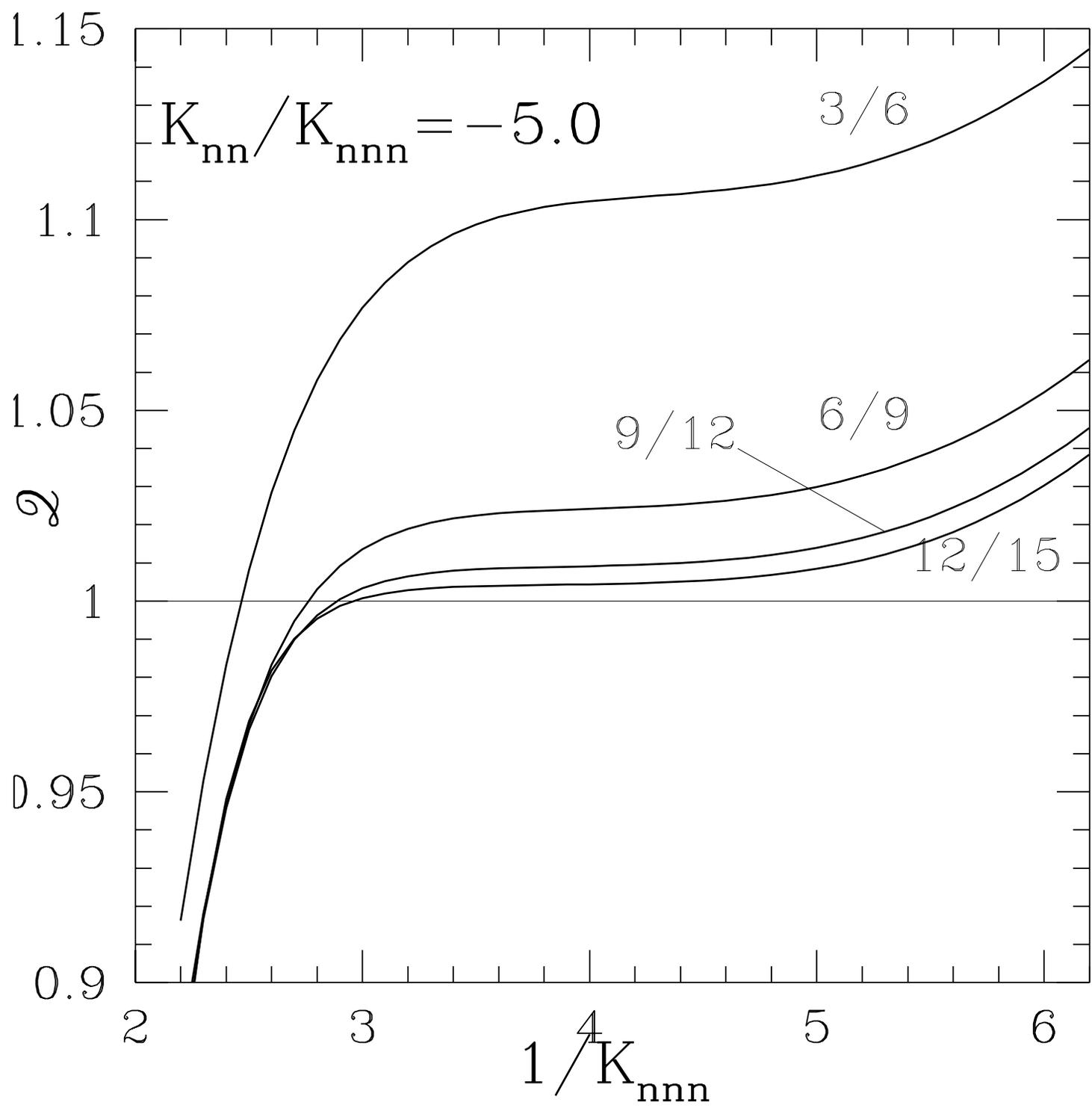
\end{document}